\documentclass{gretsi}

\usepackage[T1]{fontenc}
\usepackage[latin9]{inputenc}

\usepackage[english,french]{babel} 
\usepackage{amsmath,amssymb,amsfonts,epsfig,epstopdf,float}

\newtheorem{theorem}{ThÈorËme}
\newtheorem{definition}[theorem]{DÈfinition}

\titre{Jeux stochastiques et contrÙle de puissance distribuÈ}

\auteur{\coord{FranÁois}{MÈriaux}{1},
        \coord{MaÎl}{Le Treust}{1},
    \coord{Samson}{Lasaulce}{1},
    \coord{Michel}{Kieffer}{12}}

\adresse{\affil{1}{L2S - CNRS - SUPELEC - UniversitÈ Paris-Sud\\
    3 rue Joliot-Curie F-91192 Gif-sur-Yvette, France}
         \affil{2}{LTCI - CNRS - Telecom ParisTech\\
       46 rue Barrault F-75013 Paris, France}}

\email{francois.meriaux@lss.supelec.fr,
mael.letreust@lss.supelec.fr\\
samson.lasaulce@lss.supelec.fr, michel.kieffer@lss.supelec.fr}

\resumefrancais{Les Èmetteurs d'un canal ‡ accËs multiple sont supposÈs
choisir eux-mÍmes leur stratÈgie de contrÙle de puissance de maniËre
‡ Ítre efficaces ÈnergÈtiquement. Nous montrons que le concept de jeux stochastiques
permet de concevoir des stratÈgies de contrÙle ‡ la fois
distribuÈes, efficaces globalement et ne nÈcessitant qu'une connaissance
partielle du systËme de communication. La rÈgion de tous les points d'utilitÈ d'Èquilibre est
Ètablie et une stratÈgie pratique de contrÙle de puissance de
l'Èmetteur, reposant sur le partage temporel lÈgitime, est
proposÈe.}

\resumeanglais{Transmitters of a multiple access channel are assumed to freely choose their power control strategy in order to be energy-efficient. We show that in a stochastic game framework, we can develop energy-efficient distributed control strategies which only require partial knowledge of the entire system. Achievable utility equilibrium region is characterized and based on time-sharing, an explicit power control strategy is proposed.}

\begin{document}

\maketitle

\section{Introduction}

Dans un systËme de communication sans fil o˘ plusieurs Èmetteurs
voient leur signaux interfÈrer en rÈception, la disparitÈ des
dynamiques de puissance des composantes du signal reÁu pose
gÈnÈralement problËme au rÈcepteur. Et ce, notamment lorsque le
rÈcepteur doit dÈcoder plusieurs de ces composantes. Le contrÙle de
puissance ‡ l'Èmission vise prÈcisÈment ‡ compenser cette forte
disparitÈ. Dans cet article, nous nous intÈressons ‡ un scÈnario
d'importance croissante, celui des systËmes distribuÈs. Dans ce
cadre, l'Èmetteur dÈcide de sa politique de contrÙle de
puissance en vue de maximiser sa propre mÈtrique de performance. La
mÈtrique retenue, appelÈe utilitÈ, est l'efficacitÈ ÈnergÈtique (en
bit par Joule). Ce cadre est exactement celui introduit par Goodman
et al. dans~\cite{goodman-pc-2000}. Les auteurs de~\cite{goodman-pc-2000} ont
remarquÈ que la thÈorie des jeux, thÈorie dont l'essence mÍme est
d'Ètudier des preneurs de dÈcisions dont les actions sont
inter-dÈpendantes, est un outil pertinent pour analyser ce problËme.
Leur modËle, ‡ savoir un modËle de jeu en un coup jouÈ pour chaque
paquet de donnÈes Èmis (les joueurs Ètant les Èmetteurs et l'action
d'un joueur consistant ‡ choisir son niveau de puissance), conduit ‡
une stratÈgie de contrÙle pratique (reposant sur une connaissance
limitÈe du systËme) mais inefficace globalement. Plus prÈcisÈment,
on peut dÈmontrer qu'il existe une politique de contrÙle qui
Pareto-domine leur solution, c'est-‡-dire pour laquelle tous les
Èmetteurs font mieux en termes d'utilitÈ. Les auteurs de~\cite{LeTreustLasaulce(PowerControlRG)10}
ont dÈmontrÈ qu'un modËle de jeu rÈpÈtÈ \cite{Sorin92}
permet d'avoir une modÈlisation plus fine du problËme, modÈlisation
qui conduit ‡ des solutions plus efficaces globalement. L'idÈe
fondamentale et nouvelle en contrÙle de puissance, et que nous
exploitons dans cet article, est qu'il ne faut pas supposer le
contrÙle de puissance indÈpendant d'un paquet ‡ l'autre, et ceci mÍme
si les rÈalisations des gains des canaux sont indÈpendantes.
Un modËle de jeu dynamique tel que le jeu rÈpÈtÈ permet de tenir
compte du fait que les joueurs interagissent plusieurs fois et ceci conduit
‡ crÈer une corrÈlation entre les niveaux de puissances choisis par
un joueur au cours du temps, et nous le rÈpÈtons, mÍme pour des
canaux dits i.i.d. La contribution de cet article est de gÈnÈraliser les
travaux de~\cite{LeTreustLasaulce(PowerControlRG)10} en relaxant une hypothËse de
normalisation de l'utilitÈ individuelle par le gain de canal. Pour
faire cela, nous utilisons un modËle de jeux stochastiques~\cite{Shapley53}, ce qui nous amËne ‡ supprimer la
sous-optimalitÈ en termes de performances induite par la
normalisation nÈcessaire au modËle de jeu rÈpÈtÈ. Les travaux de~\cite{FudenbergYamamoto09,HornerSugayaTakahashiVieille09} sont alors utilisÈs pour obtenir un \textit{Folk} thÈorËme qui caractÈrise la rÈgion des utilitÈs atteignables de ce jeu stochastique. Nous prÈsentons Ègalement une stratÈgie de contrÙle de puissance explicite pour ce jeu.

Dans le paragraphe~\ref{Sec:model}, nous dÈtaillons le modËle du jeu stochastique que nous considÈrons. Au paragraphe~\ref{Sec:theory}, nous prÈsentons les rÈsultats analytiques obtenus en ce qui con- cerne la rÈgion des utilitÈs atteignables ainsi que les rÈsultats d'Èquilibre et de performance de la stratÈgie de SÈlection des Meilleurs Utilisateurs (SMU). Dans le paragraphe~\ref{Sec:simu} sont prÈsentÈs les rÈsultats de simulation obtenus pour comparer la stratÈgie SMU ‡ d'autres stratÈgies de contrÙle de puissance.

\section{ModÈlisation du problËme par un jeu stochastique}
\label{Sec:model}

Nous considÈrons un canal ‡ accËs multiple, dÈcentralisÈ au sens du
contrÙle de puissance, pour lequel $K$ utilisateurs transmettent vers un rÈcepteur sur
des intervalles de temps (durÈe d'un paquet), que nous appellerons
Ètapes du jeu rÈpÈtÈ, sur lesquels les canaux sont supposÈs
statiques. ¿ chaque Ètape, les canaux sÈlectifs en temps mais non
sÈlectifs en frÈquence, notÈs $h_i$, sont tirÈs de maniËre
indÈpendante sur un ensemble admissible~:  $|h_i|^2 \in
[\eta_i^{min},\eta_i^{max}]=\Gamma_i$. Nous supposons vÈrifiÈe l'hypothËse de
rÈciprocitÈ des canaux montants et descendants. De plus, nous
supposons que les terminaux sont capables d'estimer avec une erreur
nÈgligeable leur canaux montants (via un mÈcanisme de sÈquences
d'apprentissage, une boucle de retour, etc). Le signal reÁu peut
s'Ècrire~:
\begin{equation}
Y=\sum_{i=1}^K h_i X_i+Z
\end{equation}
avec $\mathbb{E}|X_i|^2=p_i$ et $Z\sim \mathcal{N}(0,\sigma^2)$. Dans un contexte o˘ le rÈcepteur dÈcode le signal de chaque Èmetteur sÈparÈment et o˘ il n'y a pas de mÈcanisme tel que la formation de voie~\cite{Veen-88} pour attÈnuer les interfÈrences, pour chaque utilisateur $i\in \mathcal{K}=\{1,2,...,K\}$, le rapport
signal sur interfÈrence plus bruit (RSIB) est donnÈ par :
\begin{equation}
\mathrm{RSIB}_i=\gamma_i=\frac{p_i|h_i|^2}{\sum_{j\neq i}
p_j|h_j|^2+\sigma^2}
\end{equation}
Nous pouvons maintenant dÈfinir le jeu stochastique qui
modÈlise l'interaction entre les Èmetteurs qui choisissent leur
niveau de puissance au cours du temps.


\begin{definition}[Jeu stochastique]
\label{def:stoch game}
Un jeu stochastique avec observation parfaite est dÈfini par l'uplet :

\begin{equation}
\mathcal{G}=(\mathcal{K},(\mathcal{T}_i)_{i \in \mathcal{K}},
(v_i)_{i \in \mathcal{K}},(\Gamma_i)_{i \in \mathcal{K}},\pi, \Theta),
\end{equation}
avec $\mathcal{K}$ l'ensemble des joueurs, $\mathcal{T}_i$ l'ensemble des stratÈgies pour le joueur $i$, $v_i$ la fonction d'utilitÈ du joueur $i$ sur le long terme, $\Gamma_i$ l'intervalle des Ètats de canaux accessibles au joueur $i$, $\pi$ la probabilitÈ de transition sur les Ètats et $\Theta$ l'espace des observations.
\end{definition}

La stratÈgie et l'utilitÈ sur le long terme du joueur $i$ sont dÈfinies comme suit.
\begin{definition}[StratÈgie des joueurs]
La stratÈgie du joueur $i \in \mathcal{K}$ est une sÈquence de fonctions 
$\left(\tau_{i,t} \right)_{t \geq 1}$ avec
\begin{equation}
\tau_{i,t}: \left|
\begin{array}{ccc}
\Theta^t & \rightarrow & \mathcal{A}_i \\
 \underline{h}_t & \mapsto & p_i(t).
\end{array}
\right.
\end{equation}
\end{definition}
¿ l'histoire $\underline{h}_t=(\theta(1)...,\theta(t-1), \underline{\eta}(t)) \in \Theta^t$ (observations passÈes et Ètat prÈsent), on associe une action $p_i(t) \in \mathcal{A}_i$.

La stratÈgie du joueur $i$ est notÈe $\tau_i$  et le vecteur de statÈgies $\underline{\tau} = (\tau_1, ...,
\tau_K)$ est nommÈ stratÈgie jointe. Une stratÈgie jointe $\underline{\tau}$ entraÓne une unique sÈquence d'actions  $(\underline{p}(t))_{t\geq 1}$.

\begin{definition}[UtilitÈ des joueurs]
Soit $\underline{\tau}$ une stratÈgie jointe.
L'utilitÈ du joueur $i \in \mathcal{K}$ sachant que l'Ètat initial du canal est $\underline{\eta}(1)$ est dÈfinie par
\begin{equation}
v_i(\underline{\tau}, \underline{\eta}(1)) = \sum_{t \geq 1} \lambda
(1 - \lambda)^{t-1}
\mathbb{E}_{\underline{\tau},\pi}\left[u_i(\underline{p}(t),
\underline{\eta}(t))|\underline{\eta}(1)\right]
\end{equation}
\end{definition}
avec $u_i(p_1,...,p_K)= \frac{R_if(\mathrm{RSIB}_i)}{p_i} \ [\mathrm{bit} / \mathrm{J}]$, l'utilitÈ instantanÈe du joueur $i$ telle que dÈfinie dans~\cite{goodman-pc-2000}. $R_i$ est le dÈbit d'Èmission du joueur $i$, $f$ est la fonction d'efficacitÈ, elle prend ses valeurs entre $0$ et $1$.
Le paramËtre $\lambda$ est appelÈ facteur d'escompte. Il peut Ítre interprÈtÈ comme une probabilitÈ d'arrÍt ou le fait que les joueurs apprÈcient diffÈremment leurs gains ‡ court terme et leurs gains ‡ long terme. 


\section{RÈsultats analytiques}
\label{Sec:theory}

\subsection{Folk ThÈorËme}
\begin{theorem}[Folk]
Soit F l'ensemble des utilitÈs atteignables et individuellement rationnelles. Sous l'hypothËse que les joueurs disposent du mÍme signal public, alors pour tout profil d'utilitÈ $\underline{u} \in F$, il existe $\lambda_0$ tel que pour tout $\lambda < \lambda_0$, il existe une stratÈgie d'Èquilibre public et parfait du jeu stochastique dont l'utilitÈ ‡ long terme vaut $\underline{u}\in F$.
\end{theorem}
Il faut noter qu'une telle caractÈrisation de la rÈgion d'utilitÈs atteignables est trËs puissante. En effet, la technique classique pour obtenir la rÈgion d'utilitÈs atteignables con- sisterait ‡ dÈterminer toutes les stratÈgies possibles pour les joueurs puis de calculer les utilitÈs correspondantes. Dans un jeu trËs simple o˘ chaque joueur n'aurait le choix qu'entre deux niveaux de puissance ‡ chaque Ètape, il faurait considÈrer $2^{N}$ stratÈgies possibles, avec $N$ le nombre d'Ètapes du jeu. D'aprËs \cite{Dutta95}, le \textit{Folk} thÈorËme nous autorise ‡ considÈrer uniquement les stratÈgies dites de Markov sans perte d'optimalitÈ, le nombre de stratÈgies ‡ Ètudier se rÈduit donc ‡ $2^{|\Gamma|}$ avec $|\Gamma|$ le nombre d'Ètats de canaux.

\subsection{StratÈgie de SÈlections des Meilleurs Utilisateurs}
\label{sec:best combi}
Obtenir une rÈgion d'utilitÈs atteignables est une chose, mais il reste ‡ dÈfinir formellement des stratÈgies efficaces dans cette rÈgion. C'est ce que nous proposons de faire avec l'introduction d'une stratÈgie dite de \textit{SÈlection des Meilleurs Utilisteurs}.

La stratÈgie proposÈe est basÈe sur le point de fonctionnement prÈsentÈ dans~\cite{LeTreustLasaulce(PowerControlRG)10}:
\begin{equation}
\label{eq:power-profile-op-pt} \forall i \in \mathcal{K}, \
\tilde{p_i}(t) =
\frac{\sigma^2}{\eta_i(t)}\frac{\tilde{\gamma}_K}{1-(K-1)\tilde{\gamma}_K}
\end{equation}
o˘ $\tilde{\gamma}_K$ est l'unique solution non nulle de 
\begin{equation}
x(1-(K-1) x)f'(x)-f(x)=0*.
\end{equation}

Contrairement au cas du jeu rÈpÈtÈ o˘ les gains des canaux sont constants, quand ces derniers varient ‡ chaque Ètape, la stratÈgie consistant ‡ ce que chaque joueur Èmette au point de fonctionnement (\ref{eq:power-profile-op-pt}) n'est plus optimale. Il se trouve qu'on obtient de meilleurs rÈsultats en termes de bien-Ítre social si on rÈduit l'ensemble des joueurs Èmettant au point de fonctionnement. Cette approche est intitulÈe stratÈgie de \textit{SÈlection des Meilleurs Utilisateurs}, elle est caratÈrisÈe de la maniËre suivante.


A chaque Ètape $t$ du jeu, le rÈcepteur fixe $\mathcal{K}^{'t}
\subset \mathcal{K}$, l'ensemble optimal de joueurs Èmettant au point de fonctionnement (\ref{eq:power-profile-op-pt}) pour maximiser la somme des utilitÈs instantannÈes des joueurs. Pour chaque joueur $i\in \mathcal{K}$ : 
\begin{itemize}
\item Si $i \in \mathcal{K}^{'t}$, il lui est recommandÈ d'Èmettre au point de fonctionnement (\ref{eq:power-profile-op-pt}) ‡ l'Ètape $t$. 
\item Si $i \notin \mathcal{K}^{'t}$, il lui est demandÈ de ne pas Èmettre ‡ cette Ètape. 
\end{itemize}

Il faut bien noter que le comportement des joueurs n'est pas imposÈ, le rÈcepteur envoie seulement des recommandations aux joueurs.
Pour assurer que cette stratÈgie soit un Èquilibre, un mÈcanisme de punition est Ètabli: si un joueur dÈvie de la stratÈgie, les autres joueurs jouent l'Èquilibre de Nash en un coup pour le restant du jeu.
L'Èquilibre de la stratÈgie est assurÈ si le maximum (en termes d'utilitÈ) que peut gagner un joueur en dÈviant ‡ une Ètape du jeu est infÈrieur ‡ ce qu'il va perdre en Ètant puni par les autres joueurs jusqu'‡ la fin du jeu. Nous obtenons alors la condition d'Èquilibre suivante :

\begin{theorem}[\'Equilibre de la stratÈgie]
\label{th:strat_eq} La stratÈgie SMU est un Èquilibre du jeu stochastique si $\ \forall i \in \mathcal{K}$
\begin{equation}
\lambda \leq  \frac{\mathbb{E}[u_i(\underline{p}^{smu},\underline{\eta})]-\mathbb{E}[u_i(\underline{p}^*,\underline{\eta})]}{\frac{R\eta_{\max}}{\sigma^2}\frac{f(\beta^*)}{\beta^*}+\mathbb{E}[u_i(\underline{p}^{smu},\underline{\eta})]-\mathbb{E}[u_i(\underline{p}^*,\underline{\eta})]}
\end{equation}
\end{theorem}
avec $\underline{p}^{smu}$ le profil de puissance rÈsultant de l'application de la stratÈgie SMU et $\underline{p}^*$ et $\beta^*$ respectivement le profil de puissance et le RSIB correspondant ‡ l'Èquilibre de Nash en un coup.

La complexitÈ de calcul nÈcessaire ‡ l'Èxecution de cette stratÈgie est faible puisqu'on peut prouver qu'‡ dÈbit d'Èmission Ègal, la sÈlection optimale de $k$ joueurs pour Èmettre au point de fonctionnement (\ref{eq:power-profile-op-pt}) est l'ensemble des $k$ joueurs avec les meilleurs gains de canaux. Ainsi dans un jeu ‡ $K$ joueur, le rÈcepteur doit comparer $K$ combinaisons de joueurs et non $2^K$.

\section{RÈsultats numÈriques}
\label{Sec:simu}
Pour l'obtention de rÈsultats numÈriques, nous utilisons la fonction d'efficacitÈ $f(\gamma)\,=\,e^{-\frac{a}{\gamma}}$ avec $a = 2^R-1$. Cette fonction est introduite dans~\cite{belmega-twc-2009}.


\begin{figure}[H]
\begin{center}
\includegraphics[width=1.1\columnwidth]{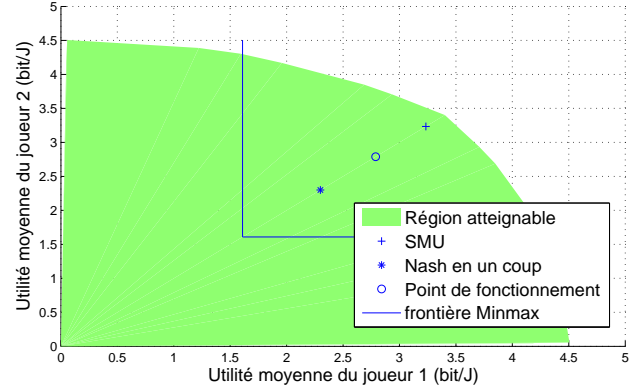}
\caption{RÈgion atteignable et utilitÈs moyennes de diverses stratÈgies pour un jeu ‡ 2 joueurs.}
\label{Fig:Folk}
\end{center}
\end{figure}

La figure~\ref{Fig:Folk} illustre la rÈgion atteignable pour un jeu ‡ $2$ joueurs et $2$ Ètats de canaux (avec $\frac{\eta_{\max}}{\eta_{\min}}=4$) en considÈrant toutes les stratÈgies possibles. La frontiËre \textit{minmax} dÈlimite la rÈgion d'Èquilibre. Les utilitÈs moyennes de SMU, du point de fonctionnement et de l'Èquilibre de Nash en un coup sont Ègalement reprÈsentÈes ‡ l'intÈrieur de cette rÈgion. Notons que que la stratÈgie SMU Pareto-domine les autres stratÈgies considÈrÈes.

%

\begin{figure}
\begin{center}
\includegraphics[width=1.1\columnwidth]{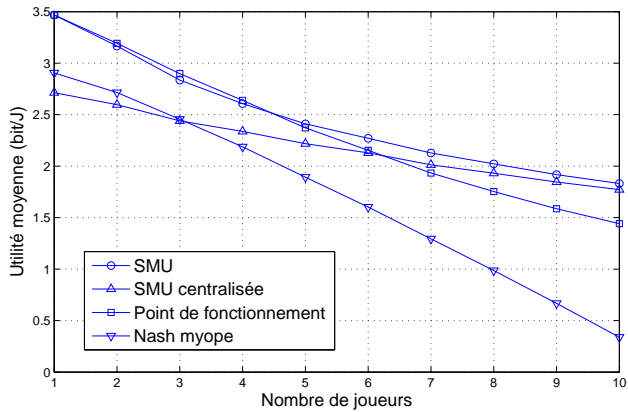}
\caption{UtilitÈs moyennes de quatre mÈcanismes de contrÙle de puissance en fonction du nombre d'Èmetteurs.}
\label{Fig:centodecen}
\end{center}
\end{figure}

La simulation prÈsentÈe en figure~\ref{Fig:centodecen} compare les utilitÈs instantannÈes moyennes de quatre mÈcanismes de contrÙle de puissance en fonction du nombre d'Èmetteurs. Pour cette simulation, on considËre un nombre fini de gains de canal. La loi d'Èvolution des gains des canaux suit la propriÈtÈ de Markov, c'est-‡-dire qu'il existe une matrice de probabilitÈ de transtion entre l'Ètat des canaux ‡ l'instant $t$ et l'Ètat des canaux ‡ l'instant $t+1$. Cette marice ainsi que les Ètats de gains de canal accessibles sont les mÍmes pour tous les joueurs.
A travers l'Ètude de ces quatres mÈcanismes, nous Ètudions les performances atteignables en fonction du caractËre centralisÈ ou dÈcentralisÈ du mÈcanisme ainsi que de la quantitÈ d'information disponible sur le systËme. Ces mÈcanismes sont les suivants :

\begin{itemize}
\item Une version centralisÈe de SMU, dans laquelle le rÈcepteur choisit qui Èmet ‡ chaque tour et impose la puissance d'Èmission en connaisant les gains des canaux ‡ l'instant $t$. Dans le modËle considÈrÈ, les Èmetteurs appliquent ‡ l'instant $t+1$ la puissance d'Èmission dÈcidÈe ‡ l'instant $t$. Ce retard se justifie par un temps de transmission entre le rÈcepteur et les Èmetteurs.
\item SMU, pour lequel le rÈcepteur dÈcide uniquement l'ensemble des Èmetteurs conseillÈ ‡ chaque tour du jeu. Chaque Èmetteur connaissant le gain de son canal et le nombres des autres Èmetteurs qui vont transmettre avec lui, il fixe lui-mÍme sa puissance d'Èmission. De la mÍme maniËre que prÈcÈdemment, on prend en compte le retard de transmission entre le rÈcepteur et les Èmeteurs. L'ensemble des joueurs qui Èmettent ‡ l'instant $t+1$ est donc dÈcidÈ par le rÈcepteur ‡ l'instant $t$.
\item La stratÈgie reposant sur le point de fonctionnement dÈveloppÈe dans~\cite{LeTreustLasaulce(PowerControlRG)10}. L'approche est encore plus dÈcentralisÈe puisque tous les Èmetteurs fixent leur puissance ‡ chaque tour en connaissant le gain de leur canal et le nombre de joueurssans recommandation de la part du rÈcepteur.
\item Un Èquilibre de Nash "myope". Dans ce cas, les Èmetteurs n'ont aucune information sur le systËme mis ‡ part l'espÈrance du gain de leur canal et le nombre de joueurs. Ils se contentent donc de jouer l'Èquilibre de Nash statique.
\end{itemize}

Il est intÈressant de noter que SMU offre de meilleures performances que les trois autres mÈcanismes. En ce qui concerne l'approche centralisÈe, le fait que la puissance d'Èmission soit connue des Èmetteurs avec un temps de retard par rapport ‡ l'Ètat des gains des canaux est un vÈritable handicap qui n'est compensÈ que pour un nombre suffisant d'Èmetteurs.

\section{Conclusion et perspectives}
Dans un rÈseau sans fil distribuÈ o˘ les Èmetteurs sont des agents Ègoistes libres de choisir leur puissance d'Èmission pour chaque paquet, les interactions ‡ long terme mÈritent d'Ítre ÈtudiÈes. Le cadre des jeux stochastiques permet de prendre en compte le caractËre rÈpÈtÈ de ces interactions ainsi que les variations des gains des canaux d'un paquet au suivant. Cette approche nous permet notamment de caractÈriser la rÈgion des utilitÈs atteignables.
Il apparaÓt qu'Ètant donnÈes les interactions sur ‡ long terme entre les Èmetteurs, ces derniers peuvent avoir intÈrÍt ‡ ne pas Èmettre certains paquets si leurs conditions de canal sont trop mauvaises. Cela nous mËne ‡ Ètablir une stratÈgie de contrÙle de puissance fondÈe sur le partage temporel qui se montre performante en termes d'efficacitÈ ÈnergÈtique.

Les perspectives de ce travail sont d'intÈgrer dans le contrÙle de puissance
plusieurs aspects visant ‡ mieux prendre en compte les
caractÈristiques des flux d'information dans des rÈseaux rÈels: la
possibilitÈ de tolÈrer un retard sur l'Èmission d'un paquet (\textit{delay
tolerant networks}); la possibilitÈ d'avoir un flux de paquets
sporadique; le fait que la taille mÈmoire de stockage des paquets ‡
l'Èmetteur est finie.
\bibliographystyle{plain}
\bibliography{BiblioMael}

\end{document}